\documentclass{aa}    

\usepackage{graphicx}
\usepackage{txfonts}
\usepackage{natbib}

\bibpunct{(}{)}{;}{a}{}{,}

\newcommand{\domdt}{\mathrm{d}\omega/\mathrm{d}t}
\newcommand{\radd}{\mathrm{rad}\,\mathrm{d}^{-2}}
\newcommand{\dadt}{\mathrm{d}a/\mathrm{d}t}

\begin{document}

 \title{Analysis of the rotation period of asteroids (1865)~Cerberus,
  (2100)~Ra-Shalom, and (3103)~Eger -- search for the YORP effect}
   
   \authorrunning{\v{D}urech et al.}
   \titlerunning{Period analysis of three near-Earth asteroids}
   
    \author{J.~\v{D}urech	\inst{1} \and
            D.~Vokrouhlick\'y	\inst{1} \and
            A.~R.~Baransky	\inst{2} \and
            S.~Breiter          \inst{3} \and
            O.~A.~Burkhonov	\inst{4} \and
            W.~Cooney 		\inst{5} \and 
            V.~Fuller           \inst{6} \and
            N.~M.~Gaftonyuk	\inst{7} \and
            J.~Gross 		\inst{5} \and
            R.~Ya.~Inasaridze   \inst{8} \and
            M.~Kaasalainen	\inst{9} \and
            Yu.~N.~Krugly	\inst{10} \and
            O.~I.~Kvaratshelia  \inst{8} \and
            E.~A.~Litvinenko    \inst{11} \and
            B.~Macomber		\inst{12} \and
            F.~Marchis		\inst{12,13} \and
            I.~E.~Molotov	\inst{14} \and
            J.~Oey		\inst{15} \and
            D.~Polishook	\inst{16} \and
            J.~Pollock		\inst{6} \and
            P.~Pravec		\inst{17} \and
            K.~S\'arneczky 	\inst{18} \and
            V.~G.~Shevchenko	\inst{7} \and
            I.~Slyusarev	\inst{7}
            R.~Stephens 	\inst{19} \and
            Gy.~Szab\'o 	\inst{18,20,23} \and
            D.~Terrell 		\inst{5} \and
            F.~Vachier		\inst{21} \and
            Z.~Vanderplate      \inst{6} \and
            M.~Viikinkoski	\inst{9} \and
            B.~D.~Warner	\inst{22} 
            }

\institute{
	Astronomical Institute, Faculty of Mathematics and Physics, Charles University,
	V Hole\v{s}ovi\v{c}k\'ach 2, 18000, Prague, Czech Republic\\
	\email{durech@sirrah.troja.mff.cuni.cz}
\and 	
	Astronomical Observatory of Taras Shevshenko National University, Astronomichna str. 3, Kiev, Ukraine 
\and    
        Astronomical Observatory Institute, Faculty of Physics, Adam
	Mickiewicz University, S{\l}oneczna 36, 60-286 Pozna\'n, Poland.
\and	Ulugh Beg Astronomical Institute,
	Uzbek Academy of Sciences, Astronomicheskaya 33, Tashkent 100052, Uzbekistan
\and	
	Sonoita Research Observatory, 1442 E Roger Rd, Tuscon, AZ 85719, USA
\and	
	Physics and Astronomy Department, Appalachian State University, Boone, NC 28608, USA        
\and	
	Crimean Astrophysical Observatory, Simeiz Department, Simeiz 98680, Ukraine
\and	
	Kharadze Abastumani Astrophysical Observatory, Ilia State University, G.Tsereteli str. 3, Tbilisi 0162, Georgia
\and	
	Department of Mathematics, Tampere University of Technology, P.O. Box 553,
        33101 Tampere, Finland
\and	
	Institute of Astronomy of Kharkiv National University, Sumska str.\ 35, 
	Kharkiv 61022, Ukraine
\and	
	The Central (Pulkovo) Astronomical Observatory of the Russian Academy of Sciences, Pulkovskoye chaussee 65/1, St.-Petersburg 196140, Russia  
\and	
	University of California at Berkeley, Department of Astronomy, 601 Campbell
        Hall, Berkeley, CA 94720, USA
\and 	
        Carl Sagan Center, SETI institute, 189 Bernardo Av., Mountain View CA 94043, USA
\and	Keldysh Institute of Applied Mathematics, RAS,
	Miusskaya sq. 4, Moscow 125047, Russia
\and	
        Kingsgrove Observatory, 23 Monaro Ave., Kingsgrove, NSW, Australia
\and	
        Department of Earth, Atmospheric, and Planetary Sciences, Massachusetts Institute of Technology,
	Cambridge, MA 02139, USA
\and	
	Ond\v{r}ejov Observatory, AV \v{C}R, 251 65 Ond\v{r}ejov, Czech Republic
\and	
	Konkoly Observatory of the Hungarian Academy of Sciences, P.O. Box 67,
        H-1525, Budapest, Hungary
\and	
	Goat Mountain Astronomical Research Station, 11355 Mount Johnson Court,
        Rancho Cucamonga, CA 91737, USA
\and    
	Department of Experimental Physics, University of Szeged, D\'om t\'er 9, Szeged H-6720, Hungary        
\and	
	Institut de M\'ecanique C\'eleste et de Calcul des \'Eph\'em\'erides, Observatoire de Paris, 
	UMR8028 CNRS, 77 Av. Denfert-Rochereau, 75014 Paris, France 
\and	
	Palmer Divide Observatory, 17955 Bakers Farm Rd., Colorado Springs, CO 80908, USA
\and    
	 ELTE Gao-Lend\"ulet Research Group, H-9700 Szombathely, Hungary
}

\offprints{J. \v{Durech}}

\date{Received ???; accepted ???}
 
\abstract{The spin state of small asteroids can change on a long timescale by the
 Yarkovsky-O'Keefe-Radzievskii-Paddack (YORP) effect, the net torque that
 arises from anisotropically scattered sunlight and proper thermal radiation
 from an irregularly-shaped asteroid. The secular change in the rotation
 period caused by the YORP effect can be detected by analysis of asteroid
 photometric lightcurves.}
{We analyzed photometric lightcurves of near-Earth asteroids (1865)~Cerberus,
 (2100)~Ra-Shalom, and (3103)~Eger with the aim to detect possible deviations from
 the constant rotation caused by the YORP effect.}
{We carried out new photometric observations of the three asteroids, combined
 the new lightcurves with archived data, and used the lightcurve inversion method
 to model the asteroid shape, pole direction, and rotation rate. The YORP effect
 was modeled as a linear change in the rotation rate in time $\domdt$. Values
 of $\domdt$ derived from observations were compared with the values predicted by
 theory.}
{We derived physical models for all three asteroids. We had to model Eger as a nonconvex body because
 the convex model failed to fit the lightcurves observed at high phase angles.
 We probably detected the acceleration
 of the rotation rate of Eger $\domdt = (1.4\pm 0.6) \times 10^{-8}\,\radd$ ($3\sigma$ error), which
 corresponds to a decrease in the rotation period by $4.2\,\mathrm{ms}\,
 \mathrm{yr}^{-1}$. The photometry of Cerberus and Ra-Shalom was consistent with a
 constant-period model, and no secular change in the spin rate was detected. We
 could only constrain maximum values of $|\domdt| < 8\times 10^{-9}\,
 \radd$ for Cerberus, and $|\domdt| < 3 \times 10^{-8}\, \radd$ for Ra-Shalom.}
{}
 
\keywords{minor planets, asteroids -- methods: data analysis --
          techniques: photometric}

\maketitle

\section{Introduction}
The anisotropic reflection of sunlight and thermal emission of an asteroid results
in a net torque that modifies the asteroid's spin state in the long term. This effect,
called Yarkovsky-O'Keefe-Radzievskii-Paddack (YORP), is important for the spin-state
evolution of asteroids in the size range from $\sim 1\,\mathrm{m}$ to $\sim 
40\,\mathrm{km}$ \citep{Bot.ea:06}. In particular, YORP can reorient the asteroid's 
spin axis, slow down or spin up their rotation, and trigger a tumbling state. YORP
has been found to be a key element in understanding the peculiar rotation pole distribution
of large Koronis-family members \citep{Vok.ea:03}, the distribution of
small members in moderately young asteroid families \citep{Vok.ea:06},
and the distribution of the rotation rate of small main-belt asteroids \citep{Pra.ea:08, Pol.Bro:09}. 
Thanks to its ability to bring asteroid rotation toward the fission
limit, it has been suggested that it is a universal formation mechanism for small binary 
asteroids \citep{Sch:07,Pra.Har:07,Wal.ea:08} and asteroid
pairs \citep{Vok.Nes:08,Pra.ea:10}. Although the 
importance of YORP is now widely recognized, the YORP effect has been detected on only three
asteroids so far: (1862)~Apollo \citep{Kaa.ea:07}, (54509)~YORP \citep{Low.ea:07,Tay.ea:07},
and (1620)~Geographos \citep{Dur.ea:08}. Here we extend this list with a new YORP 
detection for asteroid (3103)~Eger. 

The primary importance of YORP detection for specific bodies consists in
the possibility of validating the theoretical modeling of YORP itself. This
is especially important after several studies have demonstrated possible difficulties
in the theoretical prediction of the YORP effect magnitude, such as a non-uniform
distribution of density \citep{Sch.Gas:08}, a sensitive
dependence on small-scale irregularities in the shape \citep{Sta:09,Bre.ea:09} 
and the degree of thermal beaming \citep{Roz.Gre:12}. 
As a result, some cases of nondetections might be as 
important as the positive detections \citep{Sch.Gas:08,Bre.ea:09}. 
These are the cases where the simple YORP theory 
predicts a possible detection for a present data set, yet an accurate analysis
of the observations does not reveal any sign of it.

Motivated by the discussion above, we present new photometric observations of 
near-Earth asteroids (1865)~Cerberus, (2100)~Ra-Shalom, and (3108)~Eger. We 
derived a shape model and spin state for these asteroids and detected the 
acceleration of the rotation rate of Eger, which we believe is produced by the YORP
effect.%
\footnote{The whole lightcurve dataset, shape model parameters, and other
 details are available from the DAMIT website
 {\tt http://astro.troja.mff.cuni.cz/projects/asteroids3D} \citep[see also][]{Dur.ea:10}.}
For the two other asteroids we studied -- (1865)~Cerberus and
(2100)~Ra-Shalom -- we set only upper limits on the change in the rotation rate.
While in the Ra-Shalom case this observational bound is about the upper limit of the
expected theoretical YORP value for a body of its size and heliocentric orbit,
in the case of Cerberus, the constraint is much stronger. Reminiscent of the
case of (25143) Itokawa, the expected Cerberus YORP value is a factor of $\sim 3$
higher than its observed limit.

\section{Lightcurve inversion and search for the rotation-period change}
We used the lightcurve inversion method of \cite{Kaa.Tor:01} and \cite{Kaa.ea:01} to 
derive asteroids' shape, sidereal rotation period, and spin axis direction from 
the observed lightcurves. Following the method of \cite{Kaa.ea:03}, we assumed that the
rotation rate $\omega$ changes linearly with time as $\omega(t) = \omega_0 + 
\upsilon t$. Both $\omega_0$ and $\upsilon \equiv \domdt$ were free parameters of the 
optimization. We analyzed the difference (measured by the $\chi^2$ values of the 
goodness-of-fit) between a constant-period model ($\upsilon = 0$) and a model with 
$\upsilon \neq 0$. If this difference is small, the data can be fitted with the 
constant-period model and no deviation from the uniform rotation is detected. If, 
on the other hand, the fit for $\upsilon\neq 0$ is significantly better than for 
$\upsilon = 0$, the change in the rotation rate is detected.

We found a significant change in the rotation period only for asteroid Eger. For 
Cerberus and Ra-Shalom, the constant-period model fitted the 
available data well, and the free parameter $\upsilon$ did not improve the fit 
significantly. To estimate the maximum value of $|\domdt|$ that still agrees with
observations, we increased or decreased $\upsilon$ and found the value of
$\chi^2$, for which the discrepancy between the data and the model became significant. 

\subsection{(3103) Eger}
\cite{Kaa.ea:02} have derived a preliminary shape model of Eger from 13 lightcurves 
observed between 1986 and 1997. These lightcurves were obtained by \cite{Wisniewski_1987}, 
\cite{Wisniewski_1991}, \cite{Velichko_et_al_1992}, \cite{De_Sanctis_et_al_1994}, 
and \cite{Pravec_et_al_1998b}. The data were archived in the Uppsala Asteroid
Photometric Catalog (UAPC) \citep{Lag.ea:01b}. We observed Eger during its seven 
apparitions between 1997 and 2012. The aspect data for these new photometric
observations are listed in Table~\ref{table_Eger}. A detailed description of the reduction and measurement
procedures at the Wise Observatory can be found in \cite{Pol.Bro:08}.
CCD observations and data processing at Kharkiv, Simeiz, Maidanak, Lisnyky, and Kitab observatories
were done in the standard way, and the details can be found in \cite{Kru.ea:02b}.
It is particularly important that
during three apparitions -- 1997, 2007, and 2009 -- our observations were not
taken at the Earth close encounters of Eger, thus providing data at new viewing
geometries. This is because Eger is presently close to the 3/5 exterior mean motion
resonance with the Earth \citep{Mil.ea:89}, and so the viewing geometry
at close approaches repeats.
\begin{table}
\begin{center}
\begin{tabular}{cccrcrc}
\hline \hline
 Date &	$r$  & $\Delta$ & \multicolumn{1}{c}{$\alpha$} & $\lambda$ & \multicolumn{1}{c}{$\beta$}  & Obs. \\
      &	[AU] & [AU] 	& [deg]	   & [deg]     & [deg]			      &	\\ 
\hline	
1996 07 20.0  & 1.156    & 0.194  & 40.3     & 343.6     & $12.7$ &    Si \\ 
1997 02 05.0  & 1.459    & 0.483  &  9.8     & 145.9     & $11.0$ &    Kh \\
2001 06 24.9  & 1.294    & 0.422  & 41.3     & 326.1     & $26.8$ &    Si \\
2002 02 16.8  & 1.515    & 0.544  & 11.8     & 139.0     & $15.9$ &    On \\
2006 06 28.9  & 1.269    & 0.388  & 42.3     & 330.3     & $25.0$ &    Kh \\ 
2006 06 30.0  & 1.263    & 0.378  & 42.3     & 330.9     & $24.8$ &    Kh \\ 
2006 06 30.9  & 1.258    & 0.368  & 42.4     & 331.5     & $24.5$ &    Kh \\ 
2006 07 26.0  & 1.122    & 0.163  & 46.2     & 355.6     & $ 1.1$ &    Kh \\ 
2007 02 10.3  & 1.486    & 0.510  &  9.4     & 144.7     & $13.8$ &    SRO \\
2007 02 12.3  & 1.496    & 0.520  &  9.6     & 143.4     & $14.7$ &    SRO \\
2007 02 17.3  & 1.520    & 0.550  & 11.8     & 140.3     & $16.5$ &    GM \\
2007 02 19.0  & 1.527    & 0.560  & 12.7     & 139.5     & $17.0$ &    Si \\
2009 03 22.0  & 1.902    & 1.113  & 24.1     & 221.3     & $35.4$ &    W1 \\
2009 03 29.0  & 1.903    & 1.076  & 22.6     & 219.3     & $37.3$ &    W1 \\
2009 04 15.9  & 1.898    & 1.023  & 20.3     & 211.4     & $40.6$ &    Si \\
2009 05 17.9  & 1.867    & 1.076  & 25.7     & 196.9     & $38.3$ &    W1 \\
2009 05 18.8  & 1.866    & 1.080  & 25.9     & 196.6     & $38.1$ &    W1 \\
2009 05 24.8  & 1.857    & 1.107  & 27.5     & 195.3     & $36.8$ &    W1 \\
2011 06 01.0  & 1.410    & 0.688  & 42.3     & 317.0     & $26.4$ &  Ab   \\
2011 06 04.0  & 1.395    & 0.654  & 42.6     & 318.7     & $26.4$ &  On   \\
2011 06 04.0  & 1.394    & 0.654  & 42.6     & 318.7     & $26.4$ &  W2   \\
2011 06 05.0  & 1.390    & 0.644  & 42.6     & 319.2     & $26.4$ &  Li   \\
2011 06 05.0  & 1.389    & 0.643  & 42.6     & 319.2     & $26.4$ &  On   \\
2011 06 05.9  & 1.384    & 0.633  & 42.7     & 319.8     & $26.3$ &  W2   \\
2011 06 06.0  & 1.384    & 0.632  & 42.7     & 319.8     & $26.3$ &  On   \\
2011 06 09.0  & 1.369    & 0.600  & 42.9     & 321.4     & $26.2$ &  Li   \\
2011 06 10.0  & 1.363    & 0.589  & 43.0     & 322.0     & $26.2$ &  Ab   \\
2011 06 10.9  & 1.358    & 0.579  & 43.1     & 322.6     & $26.1$ &  W2   \\
2011 06 25.0  & 1.282    & 0.434  & 44.1     & 331.2     & $24.2$ &  On   \\
2011 06 27.0  & 1.272    & 0.414  & 44.2     & 332.5     & $23.7$ &  On   \\
2011 06 28.0  & 1.266    & 0.404  & 44.3     & 333.3     & $23.5$ &  On   \\
2011 07 02.3  & 1.242    & 0.363  & 44.7     & 336.4     & $22.0$ &  On   \\
2011 07 08.0  & 1.211    & 0.311  & 45.4     & 341.2     & $19.3$ &  W2   \\
2011 07 12.9  & 1.184    & 0.269  & 46.2     & 345.8     & $15.8$ &  Kh   \\
2011 07 22.9  & 1.130    & 0.196  & 50.4     & 358.4     & $ 3.7$ &  Kt   \\
2011 07 22.9  & 1.130    & 0.196  & 50.4     & 358.5     & $ 3.7$ &  Ab   \\
2011 07 23.0  & 1.129    & 0.196  & 50.5     & 358.6     & $ 3.5$ &  On   \\
2011 07 25.0  & 1.119    & 0.185  & 52.0     &   1.8     & $ 0.1$ &  Ab   \\
2011 07 27.0  & 1.108    & 0.175  & 54.0     &   5.4     & $-3.9$ &  Ab   \\
2011 07 27.3  & 1.106    & 0.173  & 54.4     &   6.1     & $-4.7$ &  On   \\
2011 08 13.4  & 1.023    & 0.171  & 82.0     &  52.4     & $-42.8$ & On    \\
2011 08 14.4  & 1.018    & 0.175  & 83.4     &  55.6     & $-44.1$ & On    \\
2011 10 25.3  & 0.953    & 0.581  & 76.6     & 147.1     & $-33.5$ & PR    \\
2011 10 26.3  & 0.956    & 0.584  & 76.2     & 147.7     & $-33.2$ & PR    \\
2011 10 27.3  & 0.959    & 0.587  & 75.8     & 148.2     & $-32.8$ & PR    \\
2011 10 29.3  & 0.966    & 0.592  & 74.9     & 149.3     & $-32.2$ & PR    \\
2011 10 30.3  & 0.970    & 0.594  & 74.5     & 149.8     & $-31.8$ & PR    \\
2011 11 01.3  & 0.977    & 0.598  & 73.7     & 150.8     & $-31.1$ & PR    \\
2011 11 02.3  & 0.981    & 0.600  & 73.3     & 151.3     & $-30.8$ & PR    \\
2011 11 03.3  & 0.985    & 0.602  & 72.9     & 151.8     & $-30.4$ & PR    \\
2011 12 09.1  & 1.160    & 0.587  & 58.1     & 164.9     & $-17.8$ & W2  \\
2011 12 30.0  & 1.274    & 0.530  & 46.1     & 166.4     & $-8.7$  & Ab   \\
2012 01 30.3  & 1.439    & 0.488  & 17.6     & 154.4     & $ 8.9$ &  PR   \\
\hline
\end{tabular}
\caption{\label{table_Eger}	
 Aspect data for new observations of Eger. The table lists the asteroid's
 distance from the Sun $r$ and from the Earth $\Delta$, the solar phase
 angle $\alpha$, the geocentric ecliptic coordinates of the asteroid
 $(\lambda, \beta)$, and the observatory (W1 -- Wise Observatory, 1\,m;
 W2 -- Wise Observatory, 46\,cm;
 Kh -- Kharkiv Observatory, 70\,cm; Si -- Simeiz, Crimean Astronomical Observatory, 1\,m; 
 GM -- Goat Mountain Astronomical Research Station, 35\,cm; SRO -- Sonoita Research Observatory, 35\,cm;
 On -- Ond\v{r}ejov Observatory, 65\,cm; Li -- Lisnyky, Kiev University Observatory, 70\,cm; Ab -- Abastumani Astrophysical Observatory, 1.25\,m;
 Ki -- Kitab Observatory, 40\,cm; PR -- PROMPT, 45\,cm).}
\end{center}
\end{table}

The photometric data set consists of 70 lightcurves covering more than twenty years. Some of them were
observed at unusually high phase angles larger than $75\deg$. The usual approach of lightcurve inversion -- 
using the convex shape model, together with the combination of Lommel-Seeliger and Lambert scattering laws -- 
did not succeed in fitting these high-phase lightcurves. The large discrepancy between the data and the 
model was an indication of a significant nonconvexity of the shape of Eger \citep{Dur.Kaa:03}. 
When we used Hapke's scattering model, the fit improved but was still not good. However, we obtained a 
good fit when modeling the shape of Eger as a general nonconvex body using the approach 
of \cite{Kaa.Vii:12}. 

\begin{figure}[t]
\resizebox{\hsize}{!}{\includegraphics{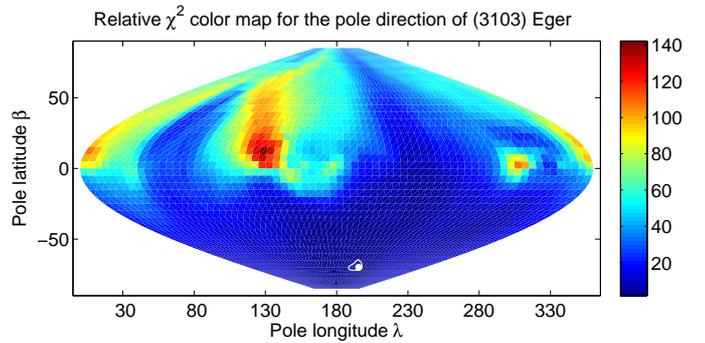}}
 \caption{\label{fig_Eger_pole}
  Statistical quality of Eger pole solutions given in sinusoidal projection
  of the sky in ecliptic coordinates. The grade of shading and the scale bar on the
  right indicate the $\chi^2$ of the fit divided by the number of data points.
  The formally best-fit case with $\lambda = 226^\circ$ and $\beta =
  -70^\circ$ is marked with a full circle. The solid line shows the contour with the $\chi^2$ value 6\%
  higher than the minimum value of the best-fit solution. It represents our region of admissible
  solutions.}
\end{figure}

We derived a unique 
solution for the sidereal rotation period $P$, the change in the rotation rate 
$\upsilon$, the pole direction $\lambda, \beta$ in ecliptic coordinates, and the shape. 
The best-fit parameters are 
as follows: $\lambda = 226^\circ \pm 15^\circ$, $\beta = -70^\circ \pm 4^\circ$,
$P = 5.710156\pm 0.000007\,$h (for JD 2446617.0), and $\upsilon = (1.4\pm 0.6)\times 
10^{-8}\,\radd$. 
These parameters and their uncertainties were derived using the weights of individual lightcurves 
that corresponded to the inverse of the noise. When all lightcurves have the same weight regardless of their quality,
the formal best-fit value of $\upsilon$ is $1.2 \times 10^{-8}\,\radd$.
The fit to the lightcurves for $\upsilon = 0$ is very good with only small phase 
shifts between the model and the data. However, using nonzero values for the $\upsilon$ parameter
improves the fit significantly. The uncertainty intervals were estimated from the increase in $\chi^2$ when varying 
the model parameters. The reported errors correspond to $3\sigma$ for the $\chi^2$ distribution with
$\sim 4700$ degrees of freedom.\footnote{\label{chisq_footnote}The $\chi^2$ distribution with $\nu$ degrees of freedom has
mean $\nu$ and variance $2\nu$. For Eger, the number of data points was $\simeq 4800$, the number
of model parameters $\simeq 100$, thus the number of degrees of freedom was $\nu \simeq 4700$.} 
Although the relative uncertainty of $\upsilon$ is large, the zero value 
is outside the formal uncertainty interval. For $\upsilon = 0$, the fit is significantly worse 
(30\% higher $\chi^2$) than for the best model. We also note that the angular difference
of $\sim 55^\circ$ between our solution and the best-fit solution by Kaasalainen 
et~al. (2002) is significant and arises from the limited number of lightcurves used to derive 
the previous model. 
The pole direction is very well constrained and is not sensitive to the value of
$\upsilon$, the scattering model, or the shape parametrization. The $3\sigma$ uncertainty region
is  shown in Fig.~\ref{fig_Eger_pole}.

As an additional check of consistency, we note that our pole
solution closely matches a constraint that was suggested from the 1991 and 1996 radar observations
\citep{Ben.ea:97}. In particular, the viewing geometry at the end of 
July 1996 was nearly equatorial for our pole as suggested by this reference.
In 1991 Aug 10, the line of sight was indeed closer to the pole by $\sim 15^\circ$.
\begin{figure*}[t]
\resizebox{\hsize}{!}{\includegraphics{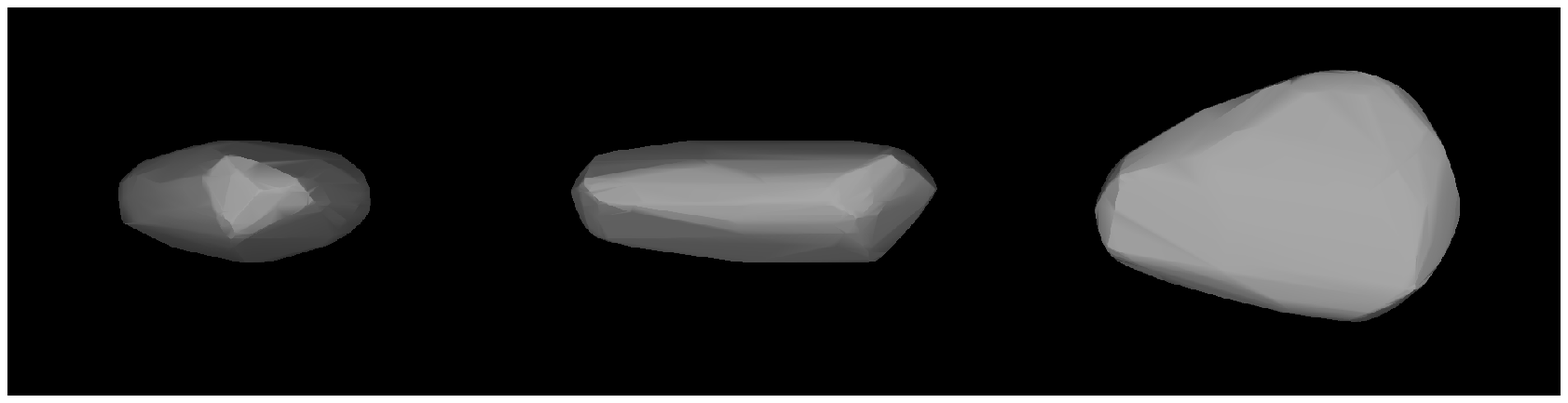}}
\resizebox{\hsize}{!}{\includegraphics{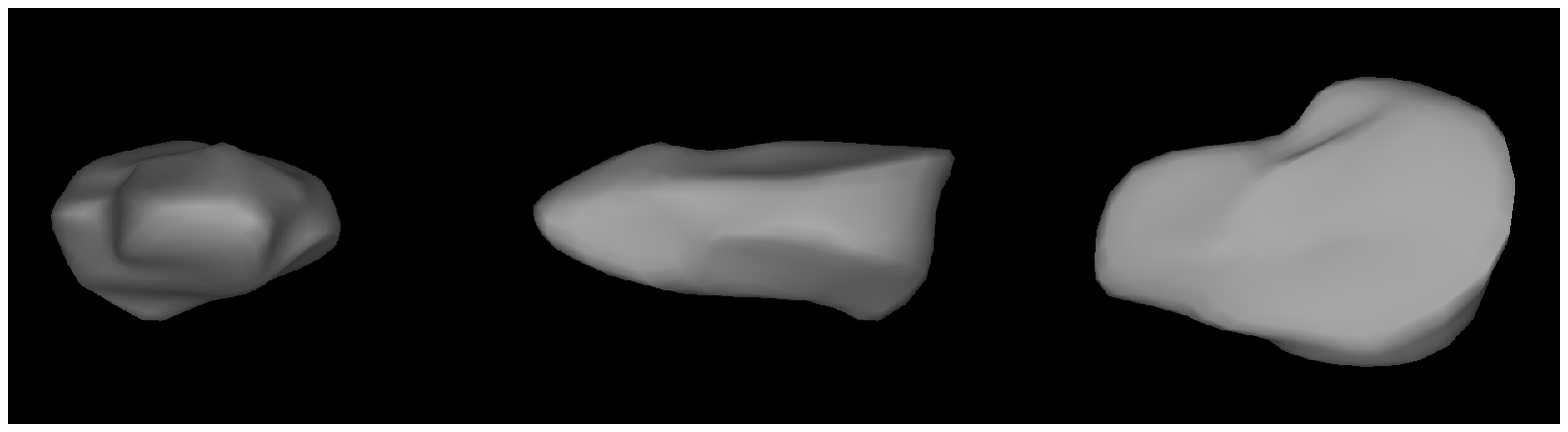}}
 \caption{\label{fig_Eger_shape}
  The convex (top) and nonconvex (bottom) shape models of Eger shown from equatorial level (left, center) and
  pole-on (right).}
\end{figure*}
\begin{figure*}[t]
 \includegraphics[width=\textwidth]{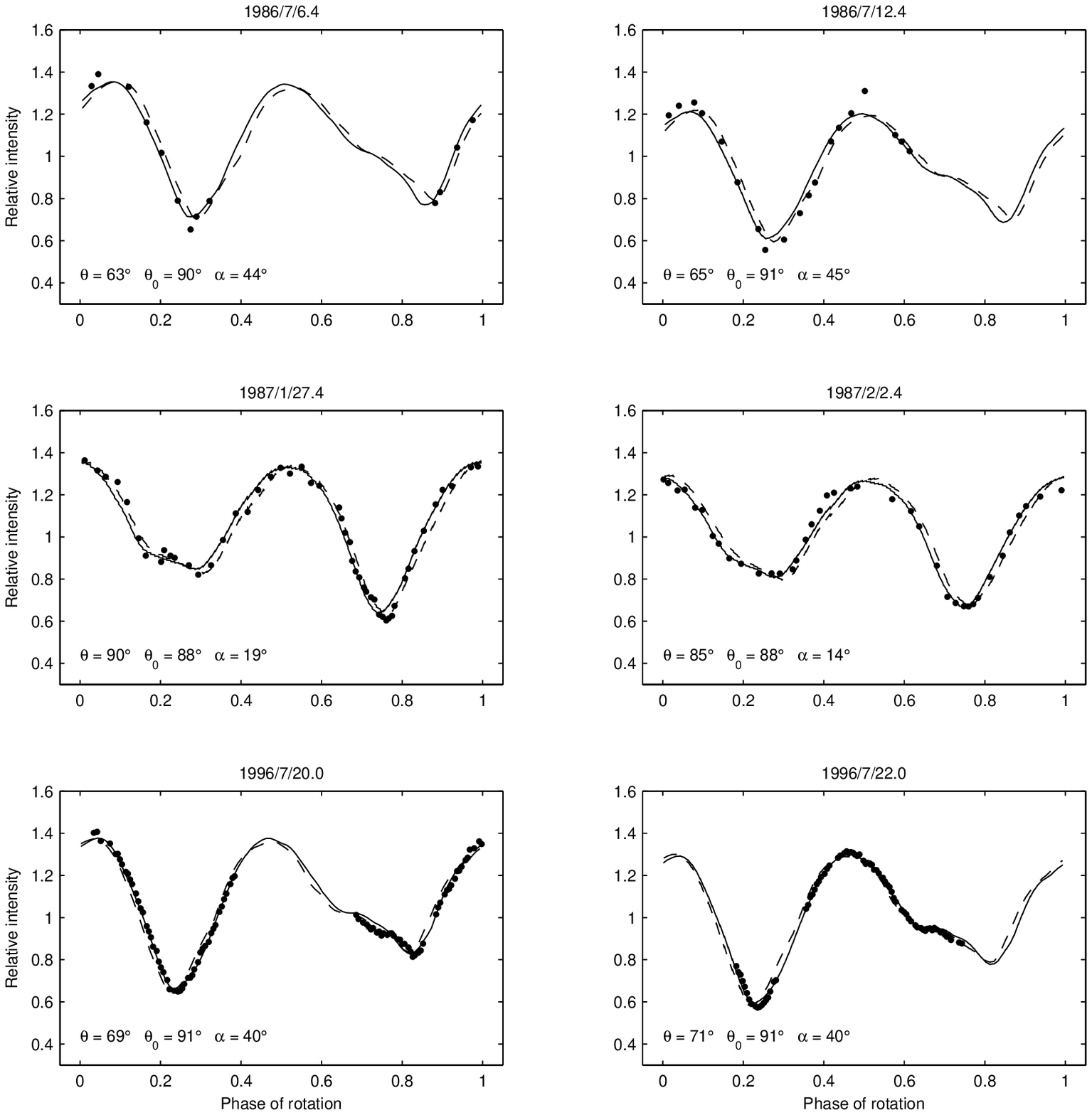}
 \caption{\label{fig_Eger_lcs}
  Examples of Eger's photometric data (points) fitted with synthetic lightcurves
  based on the convex shape model. The solid curve corresponds to the best model
  with the rotation rate accelerated by $\upsilon = 1.2 \times 10^{-8}\,\mathrm{
  rad\,d}^{-2}$, while the dashed curve corresponds to the best constant-period model
  with $\upsilon = 0$. 
  The viewing and illumination geometry is given by the aspect angle
  $\theta$, the solar aspect angle $\theta_0$, and the solar phase angle $\alpha$.}
\end{figure*}

The shape model of Eger is shown in Fig.~\ref{fig_Eger_shape}. 
Although some of the lightcurves \citep[those published by][]{Pravec_et_al_1998b}
were accurately calibrated to standard R or V magnitudes, they did not cover the viewing/illumination geometry
sufficiently to 
use them for constraining the model. Therefore, we treated all the calibrated lightcurves as relative. 

In general, nonconvex models are much less stable than convex ones with 
respect to the errors in the data, the level of regularization, the scattering model, etc. 
However, this is the first time the nonconvex shape fits the lightcurves that were observed at large phase angles 
significantly better than the convex one.
The nonconvex model in Fig.~\ref{fig_Eger_shape} is only one of many similar models that we obtained using different 
regularization, scattering models, and resolution. Apparently, the model contains many details that may lead
to misinterpretation. The main concavity is likely to be real because it repeats for models with 
different regularization. The other details differ from model to model and are instead artifacts of the 
modeling process.

\begin{figure}[t]
\resizebox{\hsize}{!}{\includegraphics{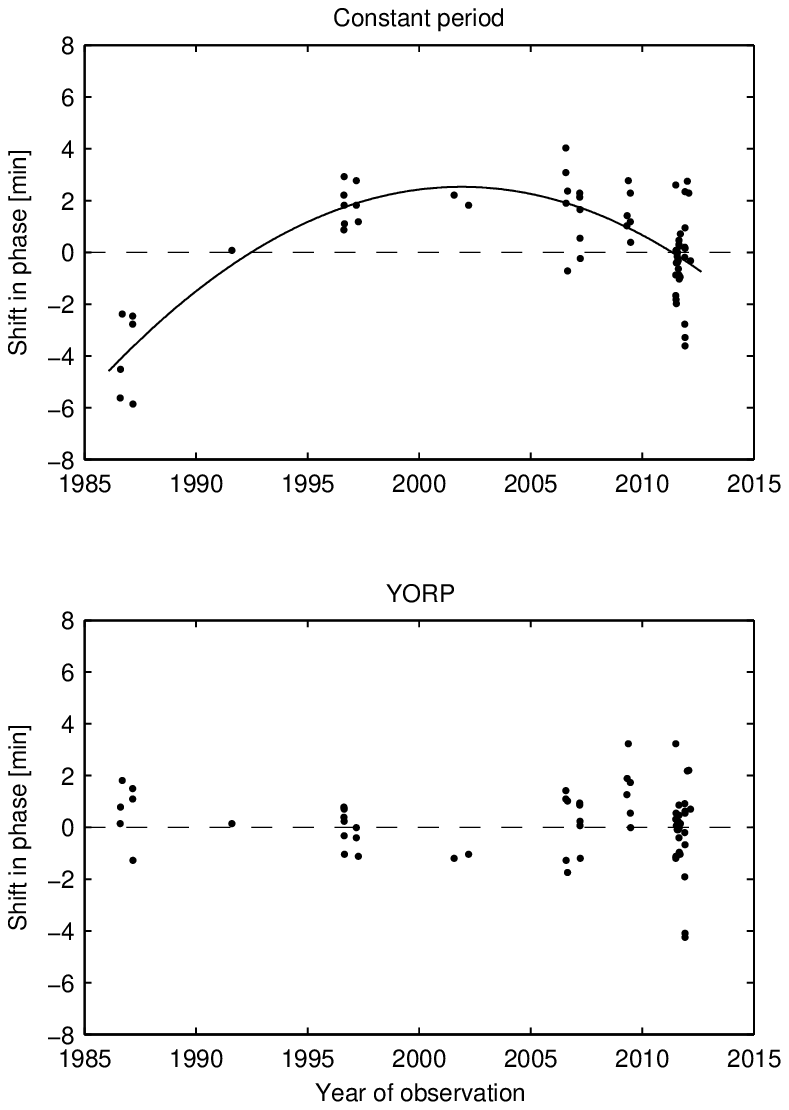}}
 \caption{\label{fig_Eger_phase}
  The shift in phase between the observed and modeled lightcurves for the constant-period model (top)
  and the model with YORP (bottom), $\upsilon = 1.2 \times 10^{-8}\,\mathrm{rad\,d}^{-2}$.
  The points in the top panel are fitted with a quadratic function.}
\end{figure}

The lightcurve fits 
for models with $\upsilon = 0$ and $\upsilon \neq 0$ are shown in Fig.~\ref{fig_Eger_lcs}.
The difference between the two models is clearly visible only for lightcurves from 1986, 1987, and
1996. Because both the number of lightcurves and the number of data points in the lightcurves increase towards 
more recent observations, there is almost no phase shift between the models for observations between 2006--2012.
The sign of the phase shift is different for different epochs. 

This is further demonstrated in 
Fig.~\ref{fig_Eger_phase}, where the shift in phase between the observed and modeled lighcurves is plotted. 
Although the scatter in the phase shifts is of minutes, there is a clear quadratic trend for the constant period model (top panel), which is what we expect if the rotation rate changes linearly in time. The best-fit quadratic
function corresponds to $\upsilon = 0.8 \times 10^{-8}\,\radd$. This is different from the value obtained 
by the lightcurve inversion, because with lightcurve inversion, the difference between the observed and modeled
brightness is minimized, not the phase shift between the observed and modeled lightcurves. In the bottom panel, the 
phase differencies are plotted for the YORP model with $\upsilon = 1.2 \times 10^{-8}\,\radd$ (equal weights for 
all lightcurves). For this value of $\upsilon$, there is no quadratic trend, and the points are randomly distributed 
along the zero value.

However, the period analysis depends critically on the observations from 1986/87 and is very sensitive
to possible systematic errors. Although the data came from three different works
\citep{Wisniewski_1987, Wisniewski_1991, De_Sanctis_et_al_1994}, 
which should largely eliminate any possible observational errors, 
there is an insufficient check against potential systematic model errors.
Therefore, our detection of the acceleration of Eger's rotation due to YORP appears plausible, 
but it will have to be nailed down with more observations in upcoming apparitions (2014, 2016, 2017, 2019, etc.) 
so that it becomes a robust detection. With more observations, the time line will be enlarged and the quadratic trend 
seen in Fig.~\ref{fig_Eger_phase} can be confirmed.

\subsection{(1865) Cerberus}
\label{sect_Cerberus_obs}
Cerberus lightcurves are characterized by unusually large amplitude, up to $2.3\,
\mathrm{mag}$, which is the largest lightcurve amplitude observed so far for any asteroid.
The UAPC contains the photometry of Cerberus from 1980 \citep{Harris_and_Young_1989}, 1989 
\citep{Wis.ea:97}, and 1998 \citep{Sar.ea:99}, while additional photometric observations
were published by \cite{Sza.ea:01}. Until now, any pole solution had not been determined.
We observed Cerberus during three apparitions in 1999, 2008, and 2009. The circumstances
of these observations are listed in Table~\ref{table_Cerberus}.
\begin{table}
\begin{center}
\begin{tabular}{cccccrcl}
\hline \hline
 Date &	$r$  & $\Delta$ & $\alpha$ & $\lambda$ & \multicolumn{1}{c}{$\beta$}  & Obs. \\
      &	[AU] & [AU] 	& [deg]	   & [deg]     & [deg]			      &	\\ 
\hline	
1999 11 03.9  & 1.432    & 0.529  & 27.3     & 359.9     & $ 4.7$ &	On       \\
1999 11 04.9  & 1.428    & 0.532  & 28.2     & 359.3     & $ 4.3$ &     On       \\
2008 09 01.9  & 1.574    & 0.822  & 34.4     &  40.4     & $14.4$ &     Ma       \\
2008 09 02.9  & 1.573    & 0.811  & 34.0     &  40.4     & $14.4$ &     Ma       \\
2008 09 24.9  & 1.535    & 0.606  & 22.6     &  35.8     & $13.4$ &     Si       \\
2008 09 30.6  & 1.521    & 0.562  & 18.1     &  33.0     & $12.7$ &     Le       \\
2008 10 20.9  & 1.455    & 0.467  &  8.5     &  17.8     & $ 7.5$ &     On       \\
2008 10 28.9  & 1.422    & 0.459  & 17.3     &  10.9     & $ 4.5$ &     Kh       \\
2008 10 29.3  & 1.420    & 0.459  & 17.8     &  10.6     & $ 4.3$ &     Li       \\
2008 11 03.8  & 1.396    & 0.463  & 24.3     &   6.3     & $ 2.1$ &     On       \\
2009 09 19.9  & 1.583    & 0.617  & 16.0     & 355.6     & $25.7$ &     Wi       \\
2009 09 20.9  & 1.583    & 0.617  & 15.9     & 354.9     & $25.5$ &     Wi       \\
2009 10 15.9  & 1.575    & 0.703  & 26.5     & 340.4     & $17.1$ &     HP       \\
2009 10 18.8  & 1.572    & 0.721  & 28.2     & 339.4     & $16.0$ &     Si       \\
\hline
\end{tabular}
\caption{\label{table_Cerberus}	
 Aspect data for new observations of Cerberus. The table lists the asteroid's
 distance from the Sun $r$ and from the Earth $\Delta$, the solar phase
 angle $\alpha$, the geocentric ecliptic coordinates of the asteroid
 $(\lambda, \beta)$, and the observatory (On -- Ond\v{r}ejov Observatory, 65\,cm;
 Kh -- Kharkiv Observatory, 70\,cm; Si -- Simeiz, Crimean Astronomical Observatory, 1\,m; 
 Le -- Leura Observatory, 36\,cm;
 Ma -- Maidanak Observatory, 1\,m; Wi -- Wise Observatory, 1\,m;
 HP -- Observatoire de Haute-Provence, 1.2\,m;
 Li -- Lick Observatory, 1\,m).}
\end{center}
\end{table}

\begin{figure*}[t]
\resizebox{\hsize}{!}{\includegraphics{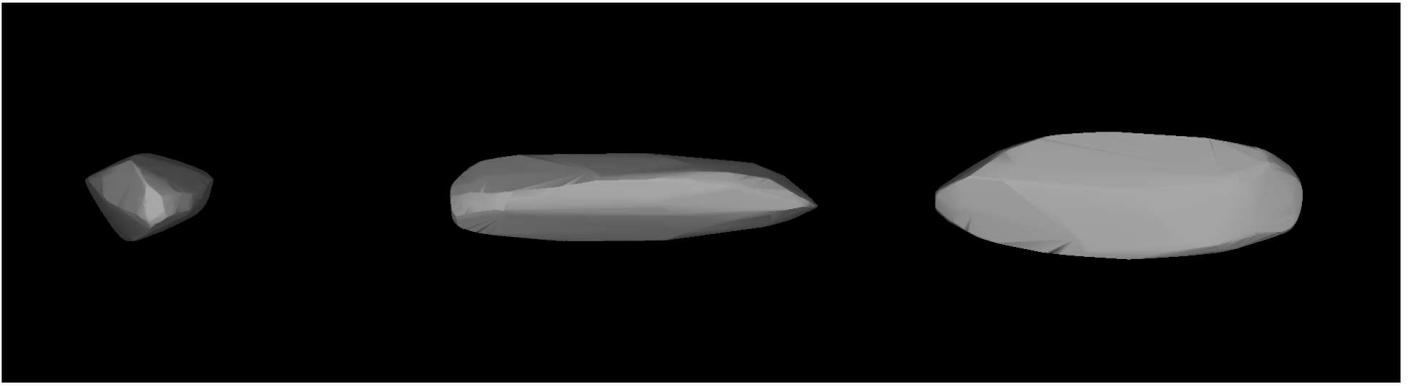}}
 \caption{\label{fig_Cerberus_shape}
  The convex shape model of Cerberus shown from equatorial level (left, center) and
  pole-on (right).}
\end{figure*}

We derived a physical model of Cerberus with the sidereal rotation period $P = 6.80328 
\pm 0.00001\,$h and the pole direction $\lambda = 298^\circ \pm 40^\circ$, $\beta = 
-72^\circ \pm 10^\circ$. 
The pole direction is well constrained, and the $3\sigma$ uncertainty region
is  shown in Fig.~\ref{fig_Cerberus_pole}.
The convex shape model (Fig.~\ref{fig_Cerberus_shape}) is very 
elongated with semiaxis ratios $a/c \simeq 4.5$ and $b/c \simeq 1.5$, suggesting that
the real shape of Cerberus might be bilobed or even consist of a close synchronous binary. 
When using calibrated lightcurves, the model is even more 
elongated and flat with $a/c \sim 7$ and $b/c \sim 2$.
The constant-period model fits well with all available lightcurves. Introducing the 
$\upsilon\neq 0$ parameter into the modeling did not improve the fit. We estimated the
maximum allowed absolute value of the period change to be $|\upsilon| < 8 \times 10^{-9}
\radd$. If the change in the rotation rate were higher, it would be detectable in the
lightcurve set as a shift between observations and the model of about $10^\circ$.
\begin{figure*}
\resizebox{\hsize}{!}{\includegraphics{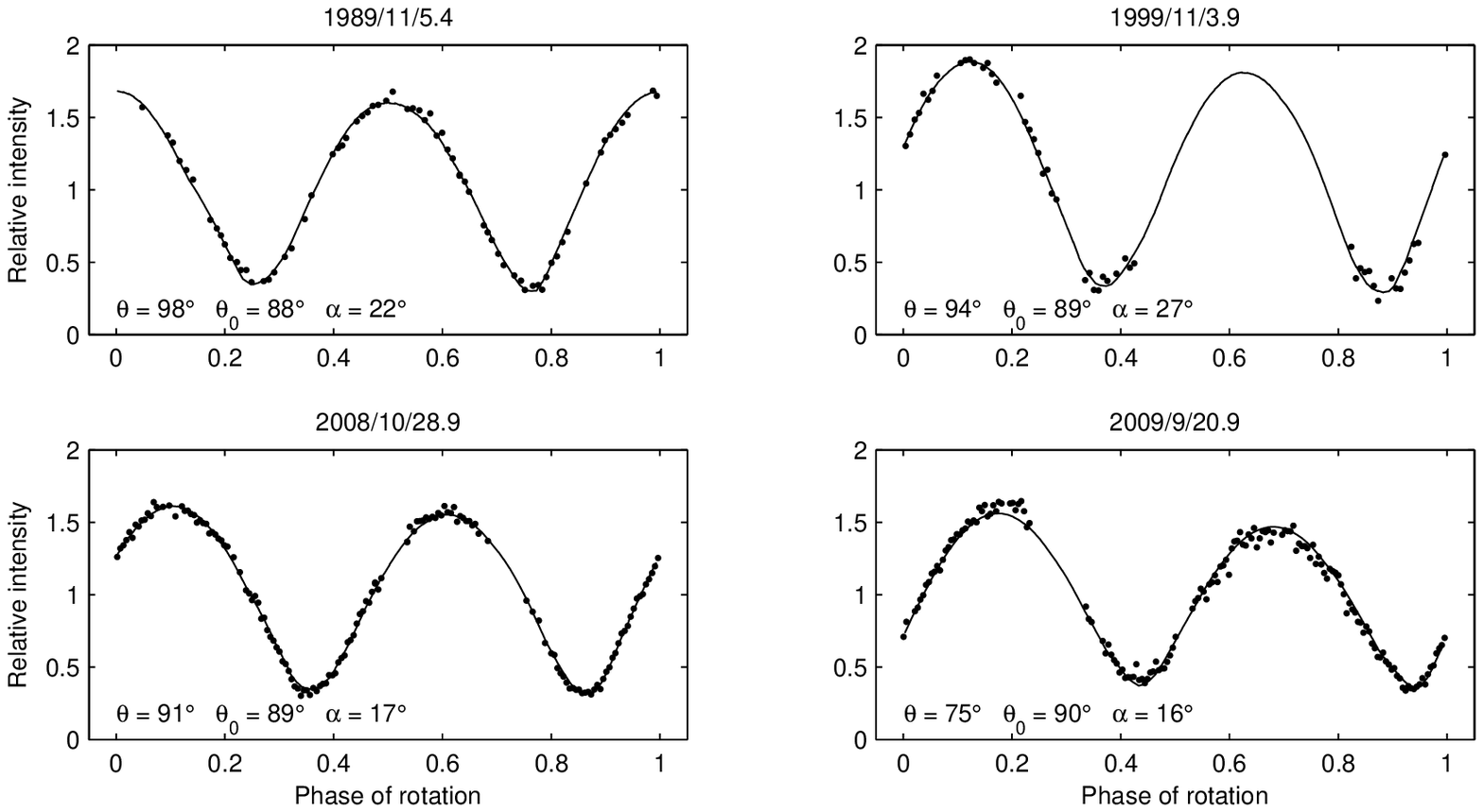}}
 \caption{\label{fig_Cerberus_lcs}
  Examples of Cerberus' lightcurves fitted with synthetic ones based on the
  convex shape model. The viewing and illumination geometry is given by the aspect angle
  $\theta$, the solar aspect angle $\theta_0$, and the solar phase angle $\alpha$.}
\end{figure*}
\begin{figure}[t]
\resizebox{\hsize}{!}{\includegraphics{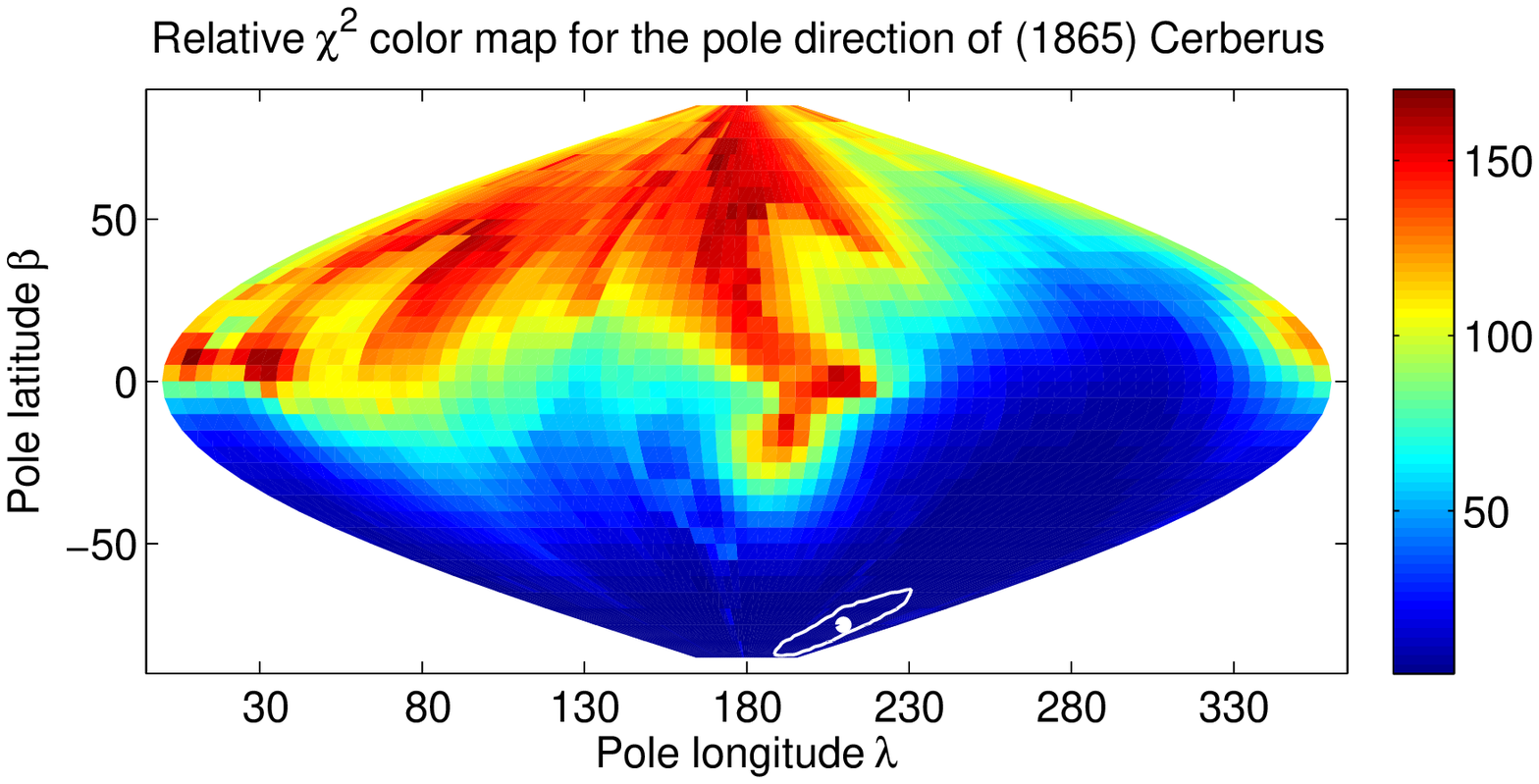}}
 \caption{\label{fig_Cerberus_pole}
  Statistical quality of Cerberus pole solutions given in sinusoidal projection
  of the sky in ecliptic coordinates. The grade of shading and the scale bar on the
  right indicate the $\chi^2$ of the fit divided by the number of data points.
  The formally best-fit case with $\lambda = 298^\circ$ and $\beta =
  -72^\circ$ is marked with a full circle. The solid line shows the contour with the $\chi^2$ value 10\%
  higher than the minimum value of the best-fit solution. It represents our region of admissible
  solutions.}
\end{figure}

\subsection{(2100) Ra-Shalom}
\label{sect_RaShalom_obs}
\begin{figure*}[t]
\resizebox{\hsize}{!}{\includegraphics{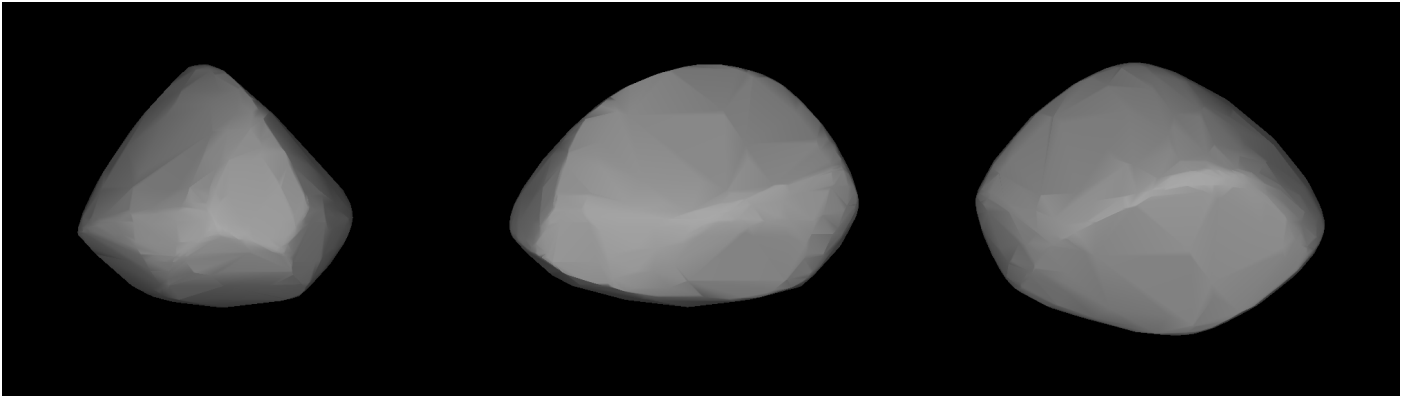}}
 \caption{\label{fig_RaShalom_shape}
  The convex shape model of Ra-Shalom shown from equatorial level (left, center) and
  pole-on (right).}
\end{figure*}

\begin{figure*}[t]
\resizebox{\hsize}{!}{\includegraphics{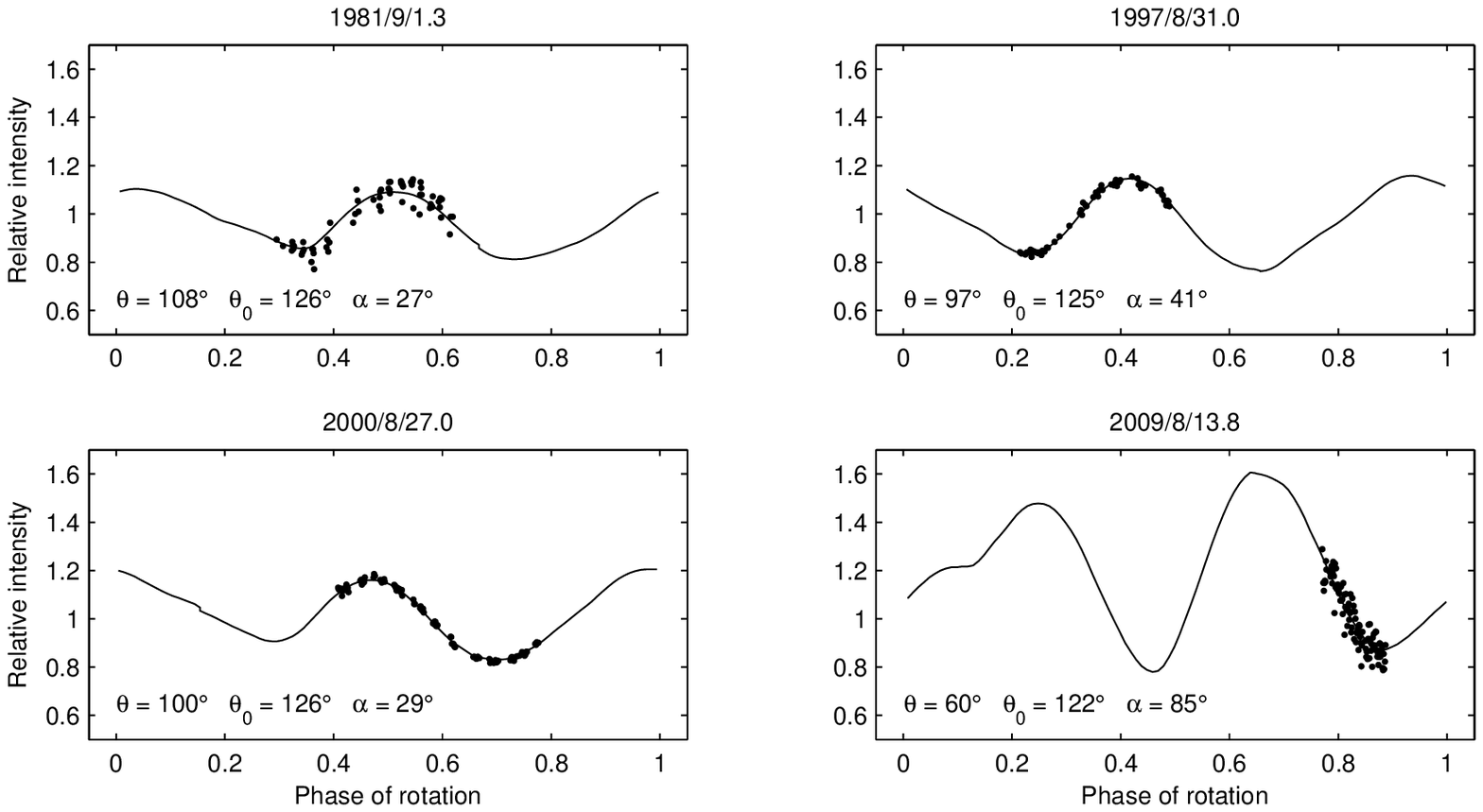}}
  \caption{\label{fig_RaShalom_lcs}
  Examples of Ra-Shalom's lightcurves fitted with synthetic ones based on the
  convex shape model. The viewing and illumination geometry is given by the aspect angle
 $\theta$, the solar aspect angle $\theta_0$, and the solar phase angle $\alpha$.}
\end{figure*}

\begin{figure}[t]
\resizebox{\hsize}{!}{\includegraphics{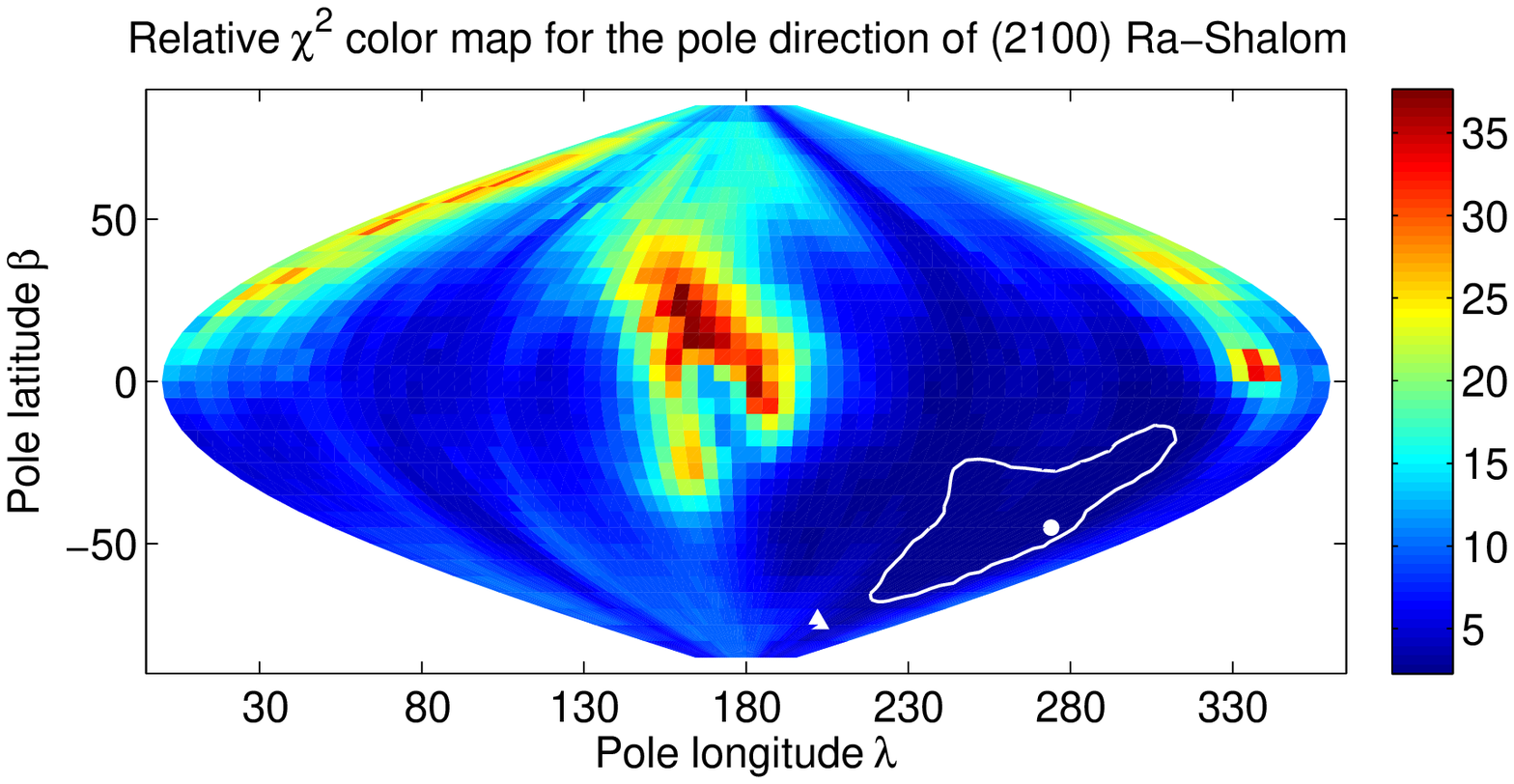}}
 \caption{\label{fig_RaShalom_pole}
  Statistical quality of Ra-Shalom pole solutions given in sinusoidal projection
  of the sky in ecliptic coordinates. The grade of shading and the scale bar on the
  right indicate the $\chi^2$ of the fit divided by the number of data points.
  The formally best-fit case with $\lambda = 313^\circ$ and $\beta =
  -45^\circ$ is marked with a full circle. The solid line shows the contour with the $\chi^2$ value 8\%
  higher than the minimum value of the best-fit solution. It represents our region of admissible
  solutions. The pole of the orbital plane corresponding to $180^\circ$ obliquity, shown
  by the triangle, is located close to this zone.}
\end{figure}

\begin{table}
\begin{center}
\begin{tabular}{cccrcrc}
\hline \hline
 Date &	$r$  & $\Delta$ & \multicolumn{1}{c}{$\alpha$} & $\lambda$ & \multicolumn{1}{c}{$\beta$}  & Obs. \\
      &	[AU] & [AU] 	& [deg]	   & [deg]     & [deg]			      &	\\ 
\hline	
2003 08 06.9  & 1.083    & 0.188  & 63.9     &   8.0     & $61.0$ &    Kh            \\
2003 08 07.0  & 1.083    & 0.188  & 63.7     &   7.7     & $61.0$ &    Kh            \\
2003 08 24.0  & 1.147    & 0.180  & 37.9     & 316.3     & $42.3$ &    Kh            \\
2003 08 25.0  & 1.150    & 0.182  & 37.1     & 314.6     & $40.6$ &    Si            \\
2003 08 26.0  & 1.153    & 0.184  & 36.5     & 313.1     & $38.9$ &    Si            \\
2003 08 27.0  & 1.156    & 0.187  & 36.0     & 311.6     & $37.2$ &    Si            \\
2003 08 27.9  & 1.158    & 0.189  & 35.7     & 310.4     & $35.6$ &    Si            \\
2003 08 28.8  & 1.160    & 0.192  & 35.4     & 309.3     & $34.1$ &    Si            \\
2003 08 29.9  & 1.163    & 0.196  & 35.3     & 308.0     & $32.3$ &    Si            \\
2003 08 30.9  & 1.166    & 0.199  & 35.3     & 307.0     & $30.7$ &    Si            \\
2003 09 01.0  & 1.168    & 0.203  & 35.5     & 305.9     & $28.9$ &    On            \\
2003 09 03.0  & 1.172    & 0.212  & 36.0     & 304.3     & $26.0$ &    On            \\
2003 09 05.9  & 1.178    & 0.225  & 37.3     & 302.4     & $21.9$ &    On            \\
2003 09 06.9  & 1.180    & 0.230  & 37.9     & 301.8     & $20.5$ &    On            \\
2003 09 14.9  & 1.190    & 0.275  & 42.8     & 299.0     & $11.8$ &    On            \\
2003 09 15.8  & 1.191    & 0.281  & 43.4     & 298.8     & $11.0$ &    On            \\
2003 09 16.8  & 1.192    & 0.287  & 44.0     & 298.7     & $10.1$ &    On            \\
2003 09 17.8  & 1.193    & 0.293  & 44.6     & 298.5     & $ 9.2$ &    On            \\
2009 08 13.8  & 0.980    & 0.362  & 84.8     & 212.2     & $33.3$ &    W2            \\
2009 08 13.8  & 0.980    & 0.362  & 84.8     & 212.2     & $33.3$ &    W1            \\
2009 08 14.8  & 0.986    & 0.363  & 83.8     & 214.4     & $33.0$ &    W2            \\
2009 08 16.8  & 0.998    & 0.365  & 81.8     & 218.9     & $32.2$ &    W1            \\
2009 08 17.8  & 1.004    & 0.367  & 80.8     & 221.0     & $31.7$ &    W1            \\
2009 08 23.8  & 1.039    & 0.383  & 75.2     & 233.1     & $28.5$ &    Kh            \\
2009 09 19.7  & 1.150    & 0.541  & 60.8     & 268.0     & $12.7$ &    W1            \\
2009 09 20.7  & 1.153    & 0.548  & 60.5     & 268.9     & $12.2$ &    W1            \\
2009 09 21.7  & 1.156    & 0.556  & 60.3     & 269.7     & $11.7$ &    W1            \\
\hline
\end{tabular}
\caption{\label{table_RaShalom}	
 Aspect data for new observations of Ra-Shalom. The table lists the asteroid's
 distance from the Sun $r$ and from the Earth $\Delta$, the solar phase
 angle $\alpha$, the geocentric ecliptic coordinates of the asteroid
 $(\lambda, \beta)$, and the observatory (Si -- Simeiz, Crimean Astronomical Observatory, 1\,m; 
 W1 -- Wise Observatory, 1\,m;
 W2 -- Wise Observatory, 46\,cm; Kh -- Kharkiv Observatory, 70\,cm;
 On -- Ond\v{r}ejov Observatory, 65\,cm).}
\end{center}
\end{table}
The UAPC contains lightcurves of Ra-Shalom from 1978 to 1997 obtained by 
\cite{Harris_et_al_1992}, \cite{Ostro_et_al_1984}, and \cite{Pravec_et_al_1998b}.
\cite{Kaa.ea:04} used this dataset, together with a few Ond\v{r}ejov
lightcurves from Ra-Shalom's apparition in 2000, to construct a pole and shape model.
Because their data set was limited, they flagged their model to be only a
preliminary attempt. Indeed, it is significantly different from the new one we 
present here, because we used additional photometry
from 2003 and 2009 listed in Table~\ref{table_RaShalom}. The derived shape model
is shown in Fig.~\ref{fig_RaShalom_shape} and the fit to selected lightcurves in 
Fig.~\ref{fig_RaShalom_lcs}. While our model is an improvement, it is still not
perfect. In particular, the pole direction is still not well constrained,
with the formally best solution at $\lambda = 313^\circ$ and $\beta = -45^\circ$. 
The statistically admissible
solutions cover an irregularly-shaped region on the sky as shown in
Fig.~\ref{fig_RaShalom_pole}. The contour corresponds to solutions with an 8\%
higher value of $\chi^2$ than the minimum. Because the number of degrees of freedom is $\nu \simeq 2300$, 
the 8\% increase corresponds to $3\sigma$ in the $\chi^2$ distribution (see the footnote on page~\pageref{chisq_footnote}).
The statistical weight of prograde-rotating spin solutions
is negligible. All solutions in our formal $3\sigma$ region correspond to a retrograde sense 
of rotation for Ra-Shalom. That our zone of admissible pole solutions
is a subset of a similar zone presented by \cite{She.ea:08} (see Fig.~7 in 
this reference), who carefully analyzed radar and other wavelength observations
of Ra-Shalom, gives us an increased confidence in our solution. The sidereal 
rotation period is $P = 19.8201\pm 0.0004\,$h. No deviation from uniform 
rotation has been detected so far. The maximum value of the change in the rotation 
rate was estimated to $-4 \times 10^{-8} < \upsilon < 2 \times 10^{-8}\,\radd$. 

\section{Discussion}
Two of our pole solutions, namely those of Eger and Cerberus, correspond
to a near-extreme obliquity value: we have $\varepsilon\simeq 176^\circ$ for Eger and
$\varepsilon\simeq 178^\circ$ for Cerberus. Both values have about a $10^\circ$ 
uncertainty in realistic terms. While the formal obliquity value of the best-fit
solution for Ra-Shalom is $\simeq 144^\circ$, Fig.~\ref{fig_RaShalom_pole} indicates
that basically all retrograde values are possible, including the value near
$180^\circ$ (note location of the orbital south pole in this figure), so at least
two, possibly even all three, of our asteroids belong to the most populated class
of near-Earth objects, those with an extremely high value of obliquity \citep{LaS.ea:04,Kry.ea:07}. 

The retrograde sense of rotation for all our targets is
consistent with an independent estimate of the Yarkovsky acceleration in their
orbits. \cite{Che.ea:08} attempted to estimate Yarkovsky effect in orbits
of all near-Earth asteroids by including a formal along-track acceleration in
their orbital fit and by estimating the related parameter $\dadt$, namely
a secular change in the orbital semimajor axis $a$ \citep[see also][]{Nug.ea:12}. 
Because the diurnal variant
of the Yarkovsky effect is dominant in the dynamics of near-Earth asteroids,
a positive/negative value for $\dadt$ value implies a prograde/retrograde
sense of rotation of the asteroid \citep[see, for example,][]{Bot.ea:06}. In 
the case of all three asteroids discussed in this paper, \cite{Che.ea:08}
give a negative best-fit value for $\dadt$, thus requiring a retrograde sense of
their rotation. This is especially important for Ra-Shalom, for which the
previous solution by \cite{Kaa.ea:04} implies prograde rotation,
while here we find a retrograde solution.

\subsection{Theoretical YORP strengths}
We now proceed with a theoretical estimation of the YORP strength for
our targets. In a sense, this is merely a consistency check that the observed
secular increase in the rotation rate $\upsilon$ (in the case of Eger) can be
interpreted as a YORP effect detection. This is because a number of parameters,
such as exact size, bulk density, and surface conductivity, are only weakly constrained,
and even knowledge of the large-scale surface-shape features may still not
be enough for precise YORP computations \citep{Sta:09,Bre.ea:09,Roz.Gre:12,Gol.Kru:12}.

\paragraph{(3103) Eger} 
We used the model of \cite{Bre.Vok:11}
to estimate the YORP value of $\upsilon$ for our shape of Eger. 
We considered parameters of the Hapke scattering model of the E-type
asteroids \citep{Bre.Vok:11} and varied
the value of geometric albedo between $0.4$ and $0.6$. This is because
the latest solutions presented by \cite{Tri.ea:10} and \cite{Har.ea:11} 
give a little larger size of $D\simeq 1.78$\,km for a lower albedo
value of $p\simeq 0.39$, while previous solutions had smaller sizes
of $D \simeq 1.5$\,km and larger albedo \citep{Ben.ea:97}. 
We also varied the bulk density
value between $2.5$ and $3$\,g/cm$^3$ and assumed surface thermal
conductivity of $0.01$\,W/m/K. However, as was shown by \cite{Bre.ea:10}, for 1D thermal models,
the resulting YORP value of $\upsilon$ does not depend on the value of conductivity. 
We scaled the lightcurve-inversion shape model (Fig.~\ref{fig_Eger_shape})
to have the same volume as a sphere of $1.78$\,km size. A
straightforward use of our model gives values for $\upsilon$ between $4.1\times
10^{-8}\,\radd$ and $8.3\times 10^{-8}\,\radd$. However, we note that the
lightcurve inversion technique does not {\it apriori} constrain the axes of the
shape model to coincide with the principal axes of the inertia tensor.
Obviously a large difference would signal a suspicious model. Assuming
a homogeneous density distribution, we find there is about a $5^\circ$
tilt between the $z$-axes of the two systems, which is too small an angle for the
correction to be determined by photometry analysis.
Assuming thus that the body rotates about its shortest axis of the inertia
tensor, and preserving the body's shape, we now obtain $\upsilon$ values 
between $4.4\times
10^{-8}\,\radd$ and $7.5\times 10^{-8}\,\radd$. Compared to the observed
value $\upsilon = (1.4 \pm 0.6)\times 10^{-8}\,\radd$, our estimates are
a factor of three to six times higher, a situation similar to the case of (54509)
YORP \citep{Low.ea:07,Tay.ea:07}. In the latter
case, the lightcurve and radar data were not able to sample signal from
about 30\% of the surface, 
which was one of the sources of the difference
between the observed and computed $\upsilon$ values. In the case of Eger,
we only have knowledge of 
the large-scale features of the asteroid shape --
the lightcurve data do not give any information about the
small-scale irregularities of the shape and the effects of thermal beaming. 
We believe this is the
main source of the difference between the observed value for $\upsilon$ and
the one computed from YORP theory. Indeed, \cite{Ben.ea:97} provide a hint  
from the analysis of the radar ranging 
to Eger that these small-scale features might actually be very
significant. 

\paragraph{(1865) Cerberus} 
The shape of Cerberus derived above
suggests that even its large-scale structures were not accurately
determined by our convex model. Obviously, we could have tried to resolve
some nonconvexities in the model, too, but the currently available 
set of the photometric observations is not large enough to derive a 
unique model. 
We thus consider our convex model as 
the current, but certainly preliminary, state-of-the-art representation of
its shape. We used the mean Hapke parameters of the S-type asteroids
in this case \citep{Bre.Vok:11} and scaled the shape model to have the same volume as a sphere with 
the diameter of $1.6\,$km \citep{Mai.ea:11}. We also varied the
geometric albedo value between $0.1$ and $0.2$. The bulk density was 
$2.5$\,g/cm$^3$.
With these values, we obtained the YORP-predicted
$\upsilon\simeq 2\times 10^{-8}\,\radd$, which was two to three times larger
than the conservative bound we obtained from the observations in
Sect.~\ref{sect_Cerberus_obs}. Obviously, this difference may be due to a number of
unconstrained factors, the inaccurate shape model first of all. The key
to further analysis of the Cerberus case consists in obtaining
additional observations, which are needed not only for possible future
detection of the YORP effect but also for improving the shape
model itself. A good
opportunity occurred in October 2016, when this asteroid could be
observed at large phase angles, and from September to November 2017.

It is interesting to note that the photometric data are consistent
with a highly nonconvex, bilobed model, but they might also be
fitted with a very close binary system. Using the analysis
performed by \cite{Bel.Sch:08} or \cite{Sch:09}, we note
there are possible stable equilibria of two ellipsoids (or an ellipsoid and
a sphere) with reasonable density values between $1$ and $3$~g/cm$^3$ 
that could match a synchronously rotating system with Cerberus' observed 
period of $\sim 6.81$~h and small separations.

\paragraph{(2100) Ra-Shalom} 
In the case of Ra-Shalom, we scaled
our shape model to be equivalent to a sphere of $\sim 2.3$\,km size
\citep{She.ea:08,Tri.ea:10,Har.ea:11} and took a geometric albedo of $0.1$ (Ra-Shalom is an S-type asteroid). 
We used a bulk density of $2$\,g/cm$^3$ and 
our best-fit formal solution for the pole. The formal
obliquity value $\varepsilon \simeq 144^\circ$ tends to minimize 
the computed effect because YORP has been found to have a node
$\upsilon\simeq 0$ near $\varepsilon \simeq 125^\circ$ for generic
shapes \citep{Cap.Vok:04,Vok.Cap:02}. With these parameter values, we obtained 
$\upsilon\simeq -6\times 10^{-8}\,\radd$, a little larger than 
its upper bound derived from the lightcurve observations 
(Sect.~\ref{sect_RaShalom_obs}). For this value, we had a negative value of 
$\upsilon=\domdt$, implying that, for a given shape model, YORP 
should decelerate the rotation rate of the body. This conclusion
would seem consistent with the rather long rotation period of 
$\simeq 19.82$\,h of this asteroid. However, we should point out
the large current uncertainty in the obliquity of the pole 
solution (Fig.~\ref{fig_RaShalom_pole}). For instance, sampling
the uncertainty region by pole positions along the meridian
of the formally best solution (i.e., keeping longitude equal to
$313^\circ$) and taking latitude values from $-10^\circ$ to
$-75^\circ$, we obtain $\upsilon$ values from $2\times 
10^{-8}\,\radd$ to $-9\times 10^{-8}\,\radd$. As expected, for 
obliquities below $\sim 125^\circ$, we obtain a positive value for $\upsilon$;
however, for obliquities greater than $\sim 125^\circ$, which
includes the formally best-fit solution, the value for $\upsilon$ is
negative. While positive $\upsilon$ values still cannot be
excluded, the majority of pole solutions in the uncertainty interval
shown in Fig.~\ref{fig_RaShalom_pole} correspond to negative
$\upsilon$ values. As a result, if carefully observed during the 
next few years (notably in late summer 2013 and 2016), Ra-Shalom might 
become the first asteroid for which YORP will be found to decrease
the rotation rate.

\begin{acknowledgements}
 We thank an anonymous referee for helpful comments that improved the final version.
 The work of J\v{D} and DV was supported by grants P209/10/0537,
 205/08/0064, and P209/12/0229 of the Czech Science Foundation and by the
 Research Program MSM0021620860 of the Ministry of Education. The work of SB
 was supported by the Polish National Science Center -- 
 grant NN 203404139. The work of MK was supported by the Academy of Finland. 
 The work of PP was supported by grant P209/12/0229 of the Czech Science Foundation.
 DP is grateful to the AXA research fund and to the continuous support from the Wise Observatory staff.
 FM was supported by the National Science Foundation under award number AAG-08077468.
 GS was supported by Hungarian OTKA Grants K-104607, MB08C 81013, a Lendulet
 Program and a Bolyai Research Fellowship of the Hungarian Academy of
 Sciences.
 
\end{acknowledgements}
 
\bibliographystyle{aa}

\newcommand{\SortNoop}[1]{}

\end{document}